\documentclass[review]{elsarticle}

\usepackage{lineno,hyperref}
\usepackage{amsmath, amssymb}
\usepackage{caption}
\usepackage{epigraph}
\usepackage{xcolor}

\modulolinenumbers[5]

\journal{Physica A}

\begin{document}

\begin{frontmatter}

\title{
	Optimal separation between vehicles for maximum flow through a light 
	signal.}

\author{J. C. Cardona}
\address{PhyL@x Physics of light, applied optics and complex system. 
	Universidad del Atl\'antico, Barranquilla, Colombia}

\begin{abstract}
        The traffic flow through a light signal is explored by using the 
        optimal velocity model and its improvement known as full 
        velocity differences model. The 
        simulations consider a single line of identical cars, equally spaced, 
        and with no obstacles after the signal crossing line. The flow 
        dependence on vehicle's characteristics, as it's length and 
        sensitivity, are studied. Also, the influence of the attitude of 
        drivers (careful or aggressive) has been used as parameters on the 
        present work.
        It was found that the optimal separation between cars, defined as the 
        distance that allows the system to carry the higher number of vehicles 
        through the signal crossing line, is independent of the 
        sensitivity of the system, but it does depend on the aggressive or 
        careful characteristic of the drivers. The optimal separation is also 
        found to be proportional to the length of the cars for a system of 
        identical vehicles.
\end{abstract}


\end{frontmatter}

    \section{Introduction}
    Traveling by car or another vehicle is a part of daily  life for modern 
    human beings. And the amount of cars running over the streets is increasing 
    fast thanks to the 
    mass production and the lowering on the buying cost. Then the free flow 
    over the streets is 
    getting more difficult each day and it became apparent the necessity of 
    finding ways to 
    control and optimize such flow. Then, understanding how 
    the traffic flow behaves is an important matter at present days, and 
    because of that, it  has been the focus of intensive research. 
    For several decades
    the main object under investigation has been the formulation of an 
    appropriated model to describe the most of the relevant observable 
    phenomena \cite{bando1995,bando1998,Davis2003,Ez-Zahraouy2003,Liu2016267,
    	Sugiyama1999,Yu2002,treiber1999,treiber2000}. Many models has been 
    proposed and characterized 
    during the last few decades, as can be revisited in an historical overview 
    of the proposed models found in \cite{vanWageningenKessels2015}.
    Some of them are macroscopic, based on the theory of fluid mechanics; some 
    others are microscopic, taken into account every vehicle and its 
    interaction with the others. 
    
    Microscopic models, specially car 
    following models are of interest due that they can simulate traffic flow 
    using rules based on human behavior (a recent review on 
    those can be found in \cite{lazar2016review}). 
    Once an acceptable model has been built, it can be used to study some of 
    the many possibles scenarios that a multi-particle problem can lead to. 
    However, on the available scientific literature,  
    most works deals with the formulation of models and there was not much work 
    on its application to an specific problem. Only recently, some of 
    the latest works has been addressed in that direction: examples of such are 
    the effect of signal on the stability of under saturated flow 
    \cite{Jiang2014}; control by to 
    kind of periodic signals \cite{Nagatani2014110}; the green wave break down 
    \cite{kerner2011,kerner2013,wang2016-2}, and recently the analysis of the 
    trip cost on a corridor with two entrance and one exit \cite{tang2017}.
    
    In this paper, the problem of finding the optimal distance between 
    consecutive cars in a line, in front of a light signal, is addressed under 
    the light of the optimal velocity (OV) and full velocity differences 
    optimal velocity (FVDOV) models. Those models are proven to successfully 
    reproduce the main features of real traffic flow 
    \cite{bando1995,Sugiyama1999}. 
    The optimal separation between cars is defined as the distance that allow 
    the maximum flow through the signal crossing line, during the green time. 
    To the best of this author knowledge, this problem has not been studied 
    anywhere.
    
    \section{The optimal velocity model.}
    The optimal velocity model was first introduced by Bando and collaborators
    \cite{bando1995,Sugiyama1999} to model and study the dynamical behavior of
    the traffic flow. The Bando model consist in a one dimensional loop
    road of length $L$, filled with $N$ identical cars.
    The $n$th driver adjust its car velocity in terms of 
    its headway  (free way ahead), so he can safely drive as fast as possible 
    while avoiding collisions. If the headway is small, the velocity must be 
    small (safe	velocity), but if the headway is large, the velocity could be 
    adjusted up to get to the maximum velocity available or desirable. The 
    dynamical equations are
    written  as:
    \begin{align}
    \dot x_n &= v, \\
    \dot v_n      &= a \left( f( \Delta x_n)  - v_n \right).
    \label{eq:dsystem}
    \end{align}
    Where the position of the $n$th car is $x_n$, the preceding car's position
    is $x_{n+1}$ and the headway is $\Delta x_n= x_{n+1} - x_n$.
    $a$ is called the sensitivity parameter, and its value determines how the 
    vehicle react to a given impulse.
    The function $f$ is the optimal velocity (OV) function, which is set to 
    be a continuous, monotonic, bounded function. Upper bound takes into account
    that the road have a maximum speed allowed and the car can only get up to a 
    finite velocity \cite{Batista2010}.
    Then $f(\Delta x\rightarrow \infty)$ is $v_{\ell}$. The
    lower bound must be set to make the car stop for
    headways lesser than some minimum distance $h_{min}>0$, so if the headway
    is near or even lesser than $h_{min}$, the acceleration must be 
    $\dot{v}<0$, stopping the car, but avoiding 
    negative velocities. A negative velocity would mean that 
    the car is reversing its direction, which is non physical.
    A typical OV function has the form:
    \begin{equation}
    f( \Delta x_n) = \frac{v_{\ell}}{1+c}        \label{eq:ovfbando}
    \left( \tanh \left(\frac{\Delta x -b}{d}\right) + c\right),
    \end{equation}
    where $d$ is a scale factor; $b$ is a safe distance configured to
    avoid collisions that represents how careful or aggressive are the drivers; 
    and $c$ is a constant in the range $[0,1]$. Note that if $c<1$, the car is 
    allowed to move in reverse direction ($\dot x <0$), which is not realistic
    due that the minimum velocity achieved by a vehicle when avoiding collision
    is zero. Consequently, in this work a value $c=1$ is used.
    Finally, $v_\ell$ is the maximum speed a vehicle can get. The maximum speed
    is determined either by the legal limits or the car capability. 
    For the closed road, it is clear that this model has a solution where all
    cars move at the same velocity $v_{op}^*$, and every car has the same 
    headway
    $h=L/N$.
    \begin{eqnarray} \label{eq:stable}
    x_n^* = nh + v_{op}^* t, &\quad
    & v^*_{op}=v_{\ell}  \frac{\tanh\left( (h -b)/d\right) + 
    	c}{1+ c}.
    \end{eqnarray}  
    
    \subsection{Model applicability and improvements.}
    The model reproduce the main features of the vehicular traffic flow, as
    the spontaneous formation of traffic jams and the stop and go waves 
    \cite{bando1995,Sugiyama1999}. However, the reaction time delay of drivers 
    is not taken into account. Bando claimed that such inclusion were not 
    significant, due that the delay time was too short \cite{bando1998}, but 
    that conclusion was controverted by further works \cite{Davis2003, 
    	zhu2008}. Anyway, several authors still consider the human time delay 
    as 
    included in the sensitivity parameter, finding the model suitable to study 
    the generalities of traffic flow. 
    
    The biggest complaint on the OV model is maybe the fact that it predicts 
    unrealistic strong variations on the velocity, when some vehicles are 
    trying to avoid collision. Addressing this issue, several variations on
    the original model has been proposed, for instance the OV with 
    decentralized feedback control \cite{Konishi2000}, the generalized OV model 
    \cite{sawada2002}, and the inclusion of the relative velocity into the model
    \cite{Yu2002,Yu2009,Liu2016267,Yu2016,Yu2016446}.  
    
    The full velocity 
    difference OV model consist in adding a new term to the 
    eq.\,\ref{eq:dsystem}, 
    proportional to the velocity difference between the current car and its 
    leader
    
    \begin{equation}\label{eq:fvdov}
    \dot v_n = a \left( f( \Delta x_n)  - v_n \right) + \lambda 
    \left( v_{n+1} - v_{n}\right).
    \end{equation}
    
    Yu and coworkers found the stability condition to be $\frac{2\beta}{a}< 
    1+2\lambda$\cite{Yu2002}. Also, they found that the inclusion of the 
    relative velocity helps the stability of the system, preventing the strong 
    unphysical variations on the velocity in near crash situation.
    
    \section{Optimization of flux through a green light traffic signal.}
    \begin{figure}[hbt] 
    	\includegraphics[scale=0.35]{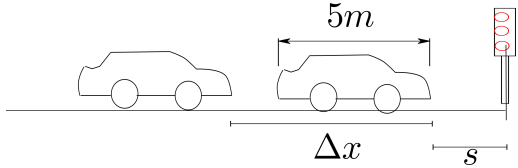}
    	\caption{\label{fig:scheme} Initial setup}
    \end{figure}
    In this work, the problem of the optimal accommodation of cars on the line 
    in front of a traffic signal is addressed. All drivers want to pass the 
    crossing line 
    before the green signal time is over, and to do that, they not only must be 
    ready to accelerate as much as they can, but also they have to take an 
    optimal	position that maximize the flux through the crossing line.
    The used model consist in a line of $N$ identical cars of  
    length $\ell$, taken as $5m$\cite{Jun2010} for the most of this work. The 
    response of the car 
    is parametrized	by the sensitivity parameter, so a 
    slow response means a small $a$ value when a large $a$ 
    represent a fast response.  In this work, this 
    parameter is taken in the range $a\in [0.2,2]s^{-1}$.
    The scale factor $d$ is taken as equal to the length of the car $\ell$. 
    The maximum velocity $v_\ell$ is set to $13.88m/s \sim 50Km/h$, which is a 
    speed typically accepted for urban area. The safe distance $d$ is a measure 
    of 
    how careful or careless (aggressive) is a driver. A system of aggressive  
    drivers will set the safety distance small, which in this work is taken as 
    $b\le 0.7\ell$, instead a system of careful drivers would set $b> 0.7\ell$. 
    For the safety distance, the used values are 
    $b\in[0.5,3]\ell$. And the green signal time duration is set in the range 
    $t\in[20,120]s$, which 
    is the usual range for urban area in Colombia.
    
    Initially, the cars are equally spaced between them, except for the first 
    one, that has its road clear (infinite headway). The first car starts
    from a distance $s=3m$ from the crossing line (see figure 
    \ref{fig:scheme}). 
    Let's call $x=0$ to the	position that mark the crossing line of the signal, 
    and the positive 
    half-axis $x>0$ after the crossing line, i.e., all cars has negative 
    initial position. The clock starts when the	signal 
    turns green and the system evolves following the dynamical rule in 
    eq.\,\ref{eq:dsystem}. After the green time is over, the cars 
    with positions $x_n>0$ have successfully passed through the signal crossing
    line.
    The position of each vehicle will be represented by a point in its front 
    (see figure \ref{fig:scheme}), so 
    a headway $\Delta x$ equal or lesser than a vehicle length means that a 
    collision has occurred, and it must be avoided.

    \section{Results and discussion.}	
    \begin{figure}[htb] 
    	\includegraphics[scale=0.5]{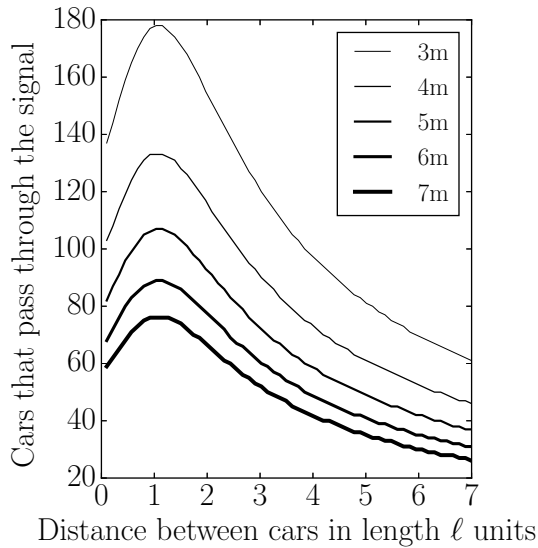}
    	\caption{\label{fig:CompareVlen} The optimal distance between cars 
    		in units of car length is always the same. Here a comparison
    		among several car lengths, for a green time signal of $110$ 
    		seconds, $a=0.2s^{-1}$ and $b=0.5\ell$
    	}
    \end{figure}
    \begin{figure*} [htb]
    	\centering
    	\includegraphics[scale=0.5]{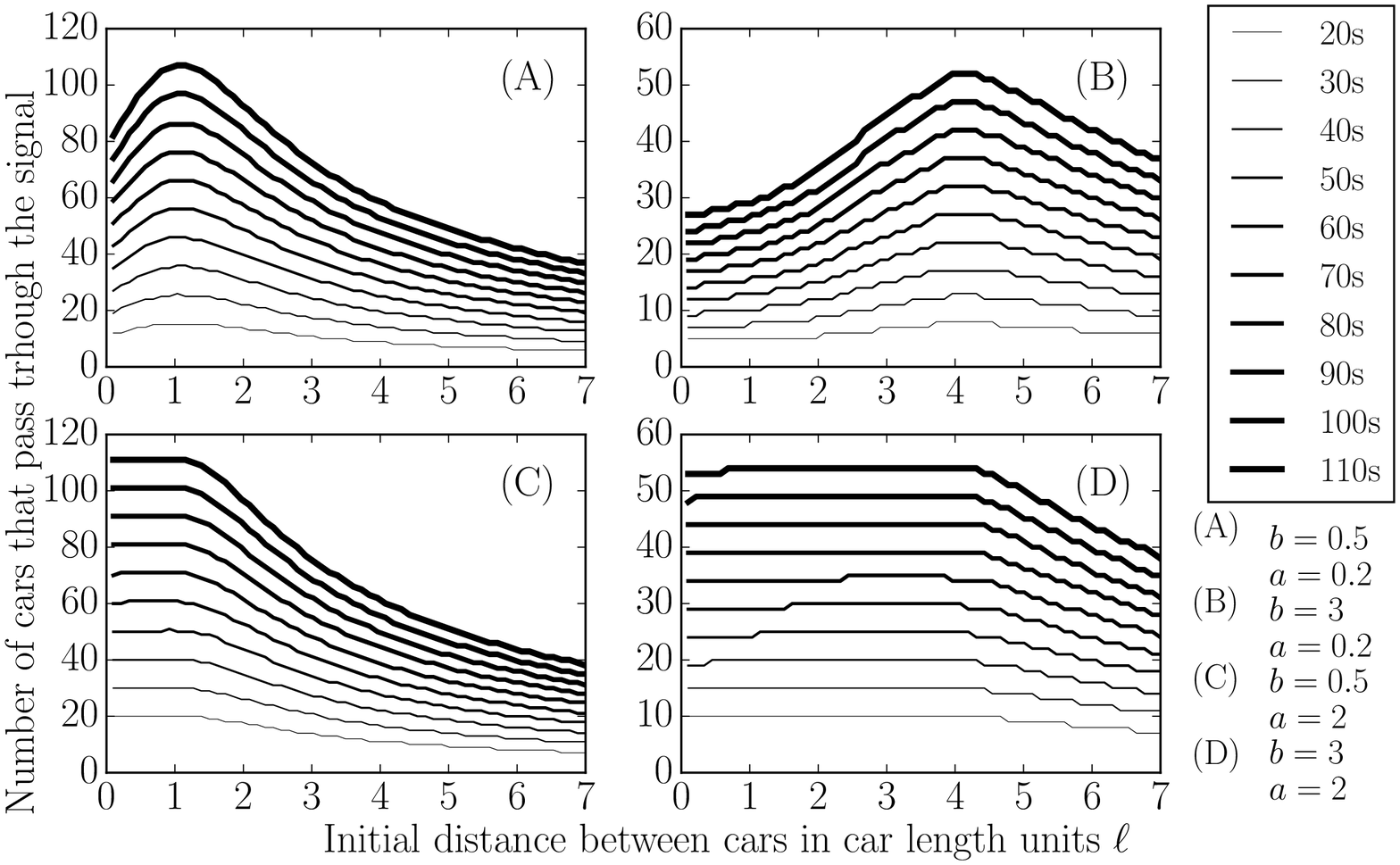}
    	\caption{\label{fig:flux} Count of cars that successfully pass through
    		the signal crossing line, for several green time durations. For all
    		computations the vehicle length is $\ell=5m$, $d=\ell$, $c=1$.  The 
    		$b$ and $a$ parameters are stated for each graph in units of $\ell$ 
    		(car lenght) and $s^{-1}$ respectively}
    \end{figure*}
    \begin{figure}[htb] 
    	\includegraphics[scale=0.5]{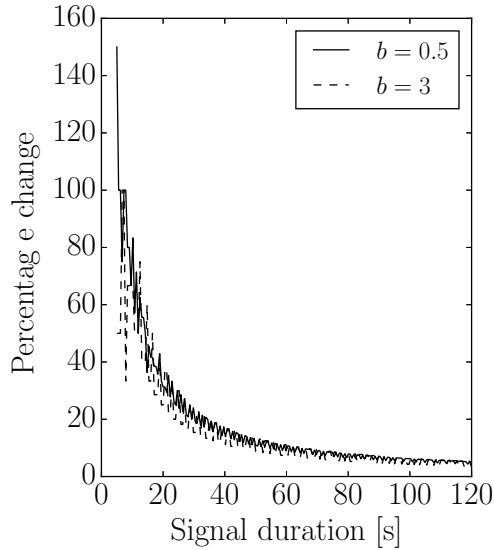}
    	\caption{\label{fig:percentage} Percentage difference between the 
    		maximum number of cars that successfully passed through the green 
    		light 
    		signal, starting form the optimal separation, when increasing the 
    		sensitivity parameter from $a=0.2s^{-1}$ to 
    		$a=2s^{-1}$. $b$ values in units of car length $\ell$}
    \end{figure}
    \begin{figure*}[htb] 
    	\includegraphics[scale=0.5]{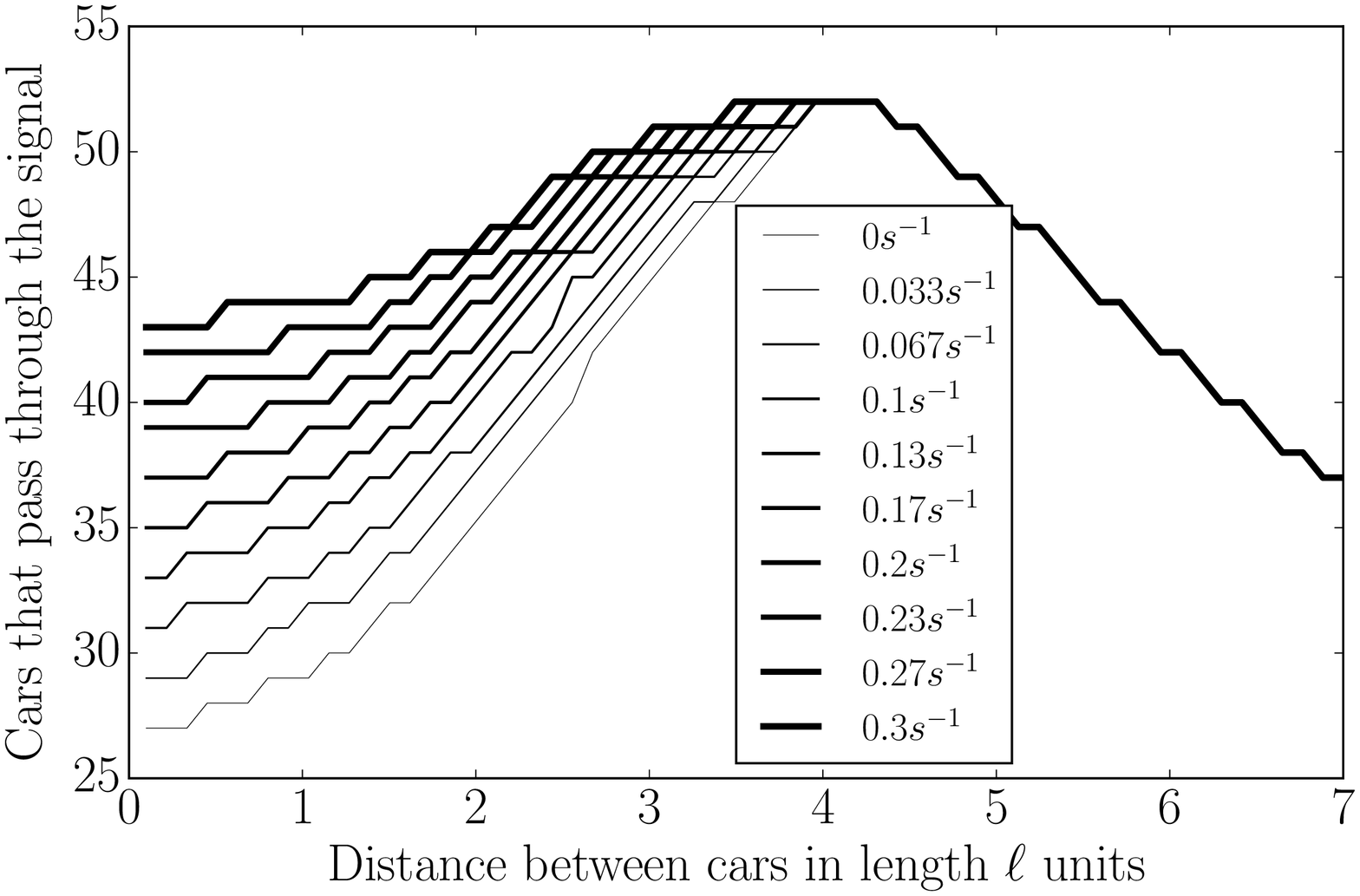}
    	\caption{\label{fig:comp_FVD} Introduction of the FVD term,  
    		for  several values of $\lambda$. The sensitivity parameter is 
    		$a=0.2s^{-1}$, $b=3\ell$, $\ell=5m$.
    	}
    \end{figure*}
    
    First, the influence of the vehicle length in the optimal 
    spacing has been tested, by holding constant the sensitivity and safety 
    parameters and 
    computing the number of vehicles that successfully pass through the signal 
    crossing line, for a given green time period. 
    The optimal separation between cars is defined as the distance that allows 
    the maximum number of cars passing through the signal crossing line for a 
    given green light time.
    In figure \ref{fig:CompareVlen} the results are 
    shown for a semaphore duration of $120s$, for several systems where the 
    only difference is the length of the vehicles. Clearly, the smaller the 
    cars are, the larger the number of cars that can successfully pass through, 
    because the
    starting position of the $n$th car is closer to the line and each system 
    is setup with the same sensitivity. Then it is expected for 
    the flux peak to be larger for smaller cars. But it comes unexpected that   
    the optimal separation, when normalized to the length of the vehicle, 
    is 
    always the same. These discovery justifies the normalization of distance to 
    the vehicle length $\ell$ on further analysis. 
    
    Next, the influence of the sensitivity and the attitude of 
    the drivers is studied.     
    The number of cars that successfully pass through the crossing line of the 
    signal, as 
    a function of the separation between cars, for several green time lapses are
    shown in figure \ref{fig:flux}. Results for  a small sensitivity parameter 
    $a=0.2s^{-1}$ are shown in sub-figures $(A)$ and $(B)$, for safety 
    distances of $b=0.5\ell$ and $b=3\ell$ respectively. And results for a 
    high sensitivity system $a=2s^{-1}$ are presented in sub figures $(C)$ and 
    $(D)$.
    It can be seen that the optimal separation between consecutive cars 
    increases with the caution of the driver. A cautious driver starts slower 
    when the leading car is close, and	only feels comfortable accelerating if 
    a 
    large enough distance is separating the cars.
    Then, in order to increase its velocity fast enough, the initial 
    separation must be large. In figure \ref{fig:flux} for a $b=3\ell$ (sub 
    figures $(B)$ and $(D)$),
    the maximum	number of passing cars is obtained around a distance of 
    $4\ell$ separating two consecutive
    cars. However, such large distance and the extreme caution of the drivers, 
    prevent the system from having a large number of successful crossings. If 
    the vehicles have a slow response, the flux is sub optimal for small 
    distances between cars, increasing strongly when the system starts near the 
    optimal separation.
    On the other hand, a set of aggressive drivers maximize the 
    passing through the signal at lesser distances (one car length is optimal 
    for $b=0.5\ell$). Surprisingly, the maximum optimal distance for maximum 
    flux shows 
    to be 
    independent of the sensitivity parameter, being dependent only on the 
    aggressive or careful behavior of the drivers. However, for large 
    sensitivities, the range of distances that allows the system to obtain the 
    maximum flux is wider.
    While the maximum flux do depend on both: the car sensitivity and the safe 
    distance; for a fixed $b$ parameter the effect of an increasing on the 
    sensitivity is an increment on the flux without changing the optimal 
    distance. Such increase is larger for 
    systems whose safety distance $b$ is farther from the region of optimal 
    flux than when the $b$ parameter is closer to the optimal value. Then, the 
    curves in figure \ref{fig:flux} are softer when the sensitivity is larger.
    For separation distances larger than the optimal value, the effect of a 
    better sensitivity is less important. To understand that, the reader must 
    remember the ``s'' shape of the hyperbolic tangent function in 
    eq.\,\ref{eq:ovfbando}; at distances between cars for which the OV 
    function is near its maximum, the nonlinearity of the hyperbolic tangent is 
    not influencing considerably the behavior of the acceleration.
    Such effect is more evident for 
    careful drivers (figure \ref{fig:flux} (B) and (D)) than for aggressive 
    ones (figure \ref{fig:flux} (A) and (C)).
    However, 
    the percentage difference between the maximum number of cars that 
    successfully passed through the green light signal,
    starting from the optimal flux separation, is approximately independent 
    from the safe distance at large signal durations (see figure 
    \ref{fig:percentage}). 
    For small signal time durations, the discrete nature of the number of cars 
    is manifested with oscillations in the percentage difference, however, a 
    careful observer would notice that the central line of those oscillation is 
    higher for aggressive drivers. Which means that for a short period of time, 
    a set of aggressive drivers is more affected by a change on the sensitivity 
    parameter than a system of careful ones, when starting from the optimal 
    distance. However, over time, the effect of safety distance is dismissed 
    and the increase on sensitivity provides the same effect on every system. 
    That can be easy understood:
    as time goes on, the vehicles are reaching its optimal 
    velocity, so the flux gets stabilized to the same number of cars by unit of 
    time, independently of the sensitivity and the safety parameters.
    
    \subsection{Computations with an improved model}
    The computations were redone under the light of the full velocity 
    differences model. The parameters used were $\lambda \in [0,0.36]$. When 
    the 
    zero value is used, the model corresponds to the original OV model of 
    eq.\,\ref{eq:dsystem}.
    In was found that the previous findings holds. The 
    optimal distance between cars remains unchanged when the $\lambda$ 
    parameter is included. The effect of the FVD term is the increasing of the 
    flux for systems starting from suboptimal separations (see figure 
    \ref{fig:comp_FVD}). Those arrangements of cars starting 
    with distance between cars equal to the optimal separation or bigger, get 
    not changed at all for the range of $\lambda$ studied.
    
    The FVDOV model takes into account not only the headway, but also the 
    relative velocity when selecting the optimal speed. Then, if the driver 
    sees that his leader is starting faster than he is, a higher acceleration 
    is possible, that makes the starting process faster and  the flux through 
    the signal for small initial distances is increased. However, the initial 
    separation between cars is determinant in the system acceleration and 
    velocity; as every vehicle but the first one starts under equal initial 
    conditions, then their acceleration is initially the same. Only the leader 
    starts without obstacles in his way, so the velocity difference between the 
    $n$th car and the $(n+1)$th is getting smaller as $n$ increase. As a 
    consequence, for initial distances equal or larger than the optimal 
    separation, the acceleration due to the OV velocity function is determinant 
    up to the free flow speed is achieved. In this process, the relative 
    velocity term is just a small perturbation and the curve in figure 
    \ref{fig:comp_FVD} is unchanged.

    \section{Conclusions}
    A comprehensive study on the effects of vehicles separation on the flux 
    through a signal light has been performed. The models used has been based 
    on the Bando's optimal velocity model, with open boundaries, and the  
    improved model known as Full Velocity Differences. The model consist on $N$ 
    identical cars, in a line, equally spaced in front of a signal, with no 
    obstacles after.
    It was found that the 
    separation between cars is indeed determinant on the flux capacity of the 
    signal. Clearly, a too large initial separation do not benefit the flux,
    but contrary to what seems to be the common thinking, a too small 
    separation prevents the system from getting 
    the maximum number of cars successfully passing through the signal. The 
    sensitivity of the systems has been found not to cause a perceptible change 
    on the flux, when the attitude of the drivers is determinant. Systems with  
    aggressive drivers are found to have small initial optimal separation. 
    Surprisingly, the 
    optimal separation between consecutive cars in terms of its length, is a 
    constant for each system, i.e., is independent of the length of the car 
    for a given configuration (parameters $a,b,c,v_\ell$ of the model). The 
    increase on the sensitivity affects the number of successfully passing 
    cars, but not the optimal initial separation. However, such increment
    is more significant when the 
    system starts from a suboptimal separation.
    It was also found that for systems starting from the optimal initial 
    separation, the percentage increment on the number of cars that cross 
    through the signal is approximately independent of the attitude for large 
    signal durations, when for short times it is larger for aggressive 
    drivers.  
    
    \section{Acknowledgement}
    Founding: This research has been supported by Universidad del Alt\'antico.
    
\section*{References}

\bibliography{bibliography}

\end{document}